%% file: main.tex
\documentclass[10pt,journal]{IEEEtran}
\usepackage{amsmath,amsfonts}
\usepackage{algorithmic}
\usepackage{array}
\usepackage{multirow}
\usepackage{textcomp}
\usepackage{stfloats}
\usepackage[noadjust]{cite}

\usepackage{url}
\usepackage{verbatim}
\usepackage{graphicx}
\usepackage{multicol}
\usepackage{booktabs}
\usepackage{subcaption}
\usepackage{lipsum}
\usepackage[dvipsnames]{xcolor}
\def\BibTeX{{\rm B\kern-.05em{\sc i\kern-.025em b}\kern-.08em
    T\kern-.1667em\lower.7ex\hbox{E}\kern-.125emX}}
\usepackage{balance}
\usepackage{comment}

\begin{document}
\title{Channel-Combination Algorithms for Robust Distant Voice Activity and Overlapped Speech Detection}
\author{Théo Mariotte, Anthony Larcher, Silvio Montrésor, Jean-Hugh Thomas\thanks{Manuscript created in September 2022. The work was conducted at LAUM and LIUM, Le Mans Université, France. This project has received funding from LMAC grant and the European Union’s Horizon 2020 research and innovation program under the Marie Sklodowska-Curie grant agreement No 101007666, the Agency is not responsible for these results or use that may be made of the information. This work was performed using HPC resources from GENCI–IDRIS (Grant 2022-AD011012565). Théo Mariotte, Jean-Hugh Thomas, and Silvio Montrésor are with the Laboratoire d'Acoustique de l'Université du Mans (LAUM), IA-GS, UMR CNRS 6613. Théo Mariotte and Anthony Larcher are with the Laboratoire d'Informatique de l'Université du Mans (LIUM), Institut Claude Chappe. Le Mans Université, 72000 Le Mans, France (e-mail: anthony.larcher@univ-lemans.fr)}}
\markboth{Accepted at IEEE/ACM Transactions on Audio Speech and Language Processing (author's version)}%
{How to Use the IEEEtran \LaTeX \ Templates}

\maketitle
\begin{abstract}
Voice Activity Detection (VAD) and Overlapped Speech Detection (OSD) are key pre-processing tasks for speaker diarization. In the meeting context, it is often easier to capture speech with a distant device. This consideration however leads to severe performance degradation. We study a unified supervised learning framework to solve distant multi-microphone joint VAD and OSD (VAD+OSD). This paper investigates various multi-channel VAD+OSD front-ends that weight and combine incoming channels.
We propose three algorithms based on the Self-Attention Channel Combinator (SACC), previously proposed in the literature. Experiments conducted on the AMI meeting corpus exhibit that channel combination approaches bring significant VAD+OSD improvements in the distant speech scenario. Specifically, we explore the use of learned complex combination weights and demonstrate the benefits of such an approach in terms of explainability. Channel combination-based VAD+OSD systems are evaluated on the final back-end task, i.e. speaker diarization, and show significant improvements. Finally, since multi-channel systems are trained given a fixed array configuration, they may fail in generalizing to other array set-ups, e.g. mismatched number of microphones. A channel-number invariant loss is proposed to learn a unique feature representation regardless of the number of available microphones. The evaluation conducted on mismatched array configurations highlights the robustness of this training strategy.
\end{abstract}


\input{01_intro.tex}

\input{02_speech_segmentation.tex}
\input{03_mch_feats.tex}
\input{04_protocol.tex}
\input{05_seg_results.tex}
\input{06_diarization}
\input{07_cSACC.tex}

\input{08_invariant.tex}

\input{09_conclu.tex}
\bibliography{habi}
\bibliographystyle{ieeetr}

\end{document}

%% file: 01_intro.tex
\section{Introduction}

\IEEEPARstart{S}{peaker} diarization answers the question \textit{Who spoke and when?} from an audio stream. 
Although much research has been conducted on speaker diarization, it remains a challenging automatic speech processing task as proven by the numerous organized challenges~\cite{ryant_third_2021,yu2022m2met}.
Most diarization systems rely on pipeline architectures based on the segmentation of the audio signal and the clustering of these segments to group them by speaker~\cite{park2022review}.
In the segmentation process, Voice Activity Detection (VAD) and Overlapped Speech Detection (OSD) are two essential sub-tasks.
VAD consists in detecting speech segments from non-speech (e.g. silence, background noise...) in an audio stream. 
In most approaches, this task is the first segmentation step in the pipeline~\cite{park2022review,bredin_pyannoteaudio_2020}.
Overlapping speech appears when several speakers are simultaneously active.
These events cause severe performance degradation in speaker diarization \cite{garcia_perera_speaker_2020}.
OSD consists of detecting segments that contain at least two simultaneous active speakers.
Then, speakers can be assigned to the overlap segments either with a heuristic approach \cite{otterson2007efficient,landini2021analysis} or by using the posterior matrix in case of Variational Bayes (VB) clustering \cite{bullock_overlap-aware_2020,landini2021analysis}.
OSD leads to significative improvements in speaker diarization \cite{garcia_perera_speaker_2020,landini2021analysis}, which makes this task essential for speaker diarization.

VAD and OSD can be processed separately in the diarization pipeline.  
They can also be extended to speaker counting~\cite{stoter2018countnet}, which detects the number of active speakers in an audio segment. 
This paper tackles joint VAD and OSD (VAD+OSD). 
It consists of the detection of non-speech ($N_{spk}=0$), speech ($N_{spk}=1$) and overlapping speech ($N_{spk}\geq2$) segments with a single model.
$N_{spk}$ is the number of active speakers.

In the multi-speaker context, such as meetings, it is often easier to record speech signals using a single distant device instead of asking each participant to carry an individual microphone.
Although a distant device offers practical benefits, the captured speech signal is more likely to be corrupted by background noise and reverberation \cite{wolfel2009distant}. 
This degradation tends to lower the performance of automatic speech processing systems~\cite{maciejewski_characterizing_2018}. 
To tackle this issue, it is a common practice to use multi-microphone devices such as microphone arrays.
These devices perform a spatial sampling that captures additional information about the spatial distribution of the sound-field~\cite{li2017multiple,he2021neural}.
The resulting signal is composed of multiple channels.
Some specific algorithms can be applied to take advantage of this spatial information implicitly contained in multi-channel signals~\cite{wolfel2009distant,benesty_microphone_2008,benesty_design_2015}.


Furthermore, the spatial distribution of the sound field is tightly related to the activity of the speakers~\cite{sivasankaran2020localization}.
The use of multi-channel audio signals, which contain such information, might be beneficial to VAD and OSD.
A few works have been conducted towards robust distant VAD and OSD~\cite{chen_mch_osd_2018,cornell_detecting_2020,Cornell2022,mariotte22_interspeech}. 
This paper investigates the use of several channel-combination algorithms as joint VAD+OSD front-ends.

\subsection{Related work}

Early studies on VAD were based on the extraction of features from the audio signal such as energy~\cite{chengalvarayan1999robust,woo2000robust}, cross-correlation~\cite{kristjansson2005voicing}, zero-crossing rate~\cite{ghaemmaghami2010noise} or linear predictive coding~\cite{nemer2001robust}.
Acoustic features have also been used to train statistical models such as Hidden Markov Models (HMM) or Gaussian Mixture Model (GMM)~\cite{sarikaya1998robust,sohn1999statistical,pfau2001multispeaker,ng2012developing}.
With the recent advances in supervised learning and deep neural networks, novel approaches have emerged. 
Early architectures were based on multi-layer perceptron (MLP)~\cite{ryant2013speech}. 
More advanced neural models such as Convolutional Neural Networks (CNN)~\cite{thomas2014analyzing} or Long Short Time Memory (LSTM) recurrent neural networks~\cite{gelly2017optimization,lavechin2019end} have then been proposed, drastically improving the VAD performance compared to signal-based or statistical approaches.

Similarly to VAD, early research on OSD mostly focuses on HMM/GMM classifiers~\cite{boakye_improved_2011,vipperla_2012_osd,charlet_impact_2013}.
Some approaches show diarization improvement by detecting and removing overlapping speech segments during the clustering step~\cite{boakye_improved_2011,sajjan2018leveraging}.
Recent OSD approaches are mostly based on deep neural networks.
LSTM neural networks were first applied to OSD in~\cite{geiger2013detecting} and have become widespread, improving the detection compared to classical machine learning approaches~\cite{sajjan2018leveraging,bullock_overlap-aware_2020}.
Alternatively, some authors use convolutional neural networks \cite{cornell_detecting_2020,kunevsova2019detection,jung21_interspeech}, such as the Temporal Convolutional Network (TCN) \cite{cornell_detecting_2020,Cornell2022}.
In~\cite{bullock_overlap-aware_2020}, Bullock~\textit{et al.} show promising diarization performance improvement by detecting overlapping segments with learnable features and a LSTM-based system.
The authors also propose an algorithm to assign the overlapping speech segments to the appropriate speaker, thus improving diarization performance.
Most VAD and OSD have a similar architecture and are formulated as binary frame classification tasks.
A few studies explore the joint training of both tasks~\cite{jung21_interspeech,cornell_detecting_2020}.
Specifically, Cornell~\textit{et al.} \cite{Cornell2022} show that both LSTM- and TCN-based architectures are suitable for joint VAD+OSD and speaker counting.


Research on VAD and OSD is mostly focused on close-talk speech signals, i.e. when the speaker is close to the microphone.
However, these conditions are rarely met in the meeting context since it is more convenient to capture the scene with a single distant device than requiring speakers to carry individual microphones. 
Few studies have been conducted on distant multichannel VAD and OSD.
In \cite{cornell_detecting_2020}, the authors propose a TCN-based architecture for distant speaker counting.
This work is extended in~\cite{Cornell2022} by adding spatial features extracted from multi-microphone signals.
The authors show a significative performance gain on both VAD and OSD by using spatial features.

Multichannel speech processing concerns various speech processing tasks such as speech recognition \cite{park2020robust,gong_self-attention_2021,watanabe2020chime}, speech separation \cite{yoshioka2018multi,gu2020enhancing}, and speaker diarization and recognition \cite{taherian2020robust,watanabe2020chime}.
The spatial information contained in these signals is often exploited through spatial filtering -- beamforming -- which consists of weighting and combining channels.
Thus, the output of the spatial filter is single-channel but steered in a given angular direction.
Early beamforming approaches are mostly based on signal processing methods~\cite{benesty_microphone_2008,sounden_optimal_2010,applebaum1976,benesty_design_2015,anguera2007acoustic} while latest algorithms exploit neural networks \cite{li2016neural,minhua2019frequency,park2020robust,gong_self-attention_2021,cornell2022learning}.
In~\cite{gong_self-attention_2021}, Gong \textit{et al.} propose a straightforward way to compute combination-weights based on a self-attention~\cite{vaswani_attention_2017} module and Short-Time Fourier Transform (STFT) magnitude features.

The performance of multi-microphone algorithms tends to highly rely on the array geometry used during training.
A few studies have been conducted on array-agnostic or channel-number-agnostic algorithms in the context of speech separation~\cite{luo2020end,taherian2020robust}.
Both approaches improve the robustness of the system to array mismatch but require more model parameters to train.


\subsection{Contributions}
Building on these previous works, this paper proposes several channel-combination algorithms as joint VAD and OSD (VAD+OSD) front-ends.
These channel combination algorithms are based on the Self Attention Channel Combinator (SACC)~\cite{gong_self-attention_2021}.
The main motivation behind this work is to include the phase information in the combination-weight estimation process.
A first extension is proposed by replacing the STFT with a learnable analytic filter bank~\cite{pariente2020filterbank} in the original SACC algorithm.
In a second approach, following the work conducted in~\cite{mariotte22_interspeech}, we explore two architectures in the complex domain: EcSACC and IcSACC. 
The former processes the STFT parts separately but doubles the number of parameters. 
The latter uses a single self-attention module to process both parts.
These extensions show similar performance as state-of-the-art approaches while being more explainable by design.
The explainability features offered by complex extensions are demonstrated by the visualization of the beampattern.
This leads to a better understanding of the use of spatial information in such systems.
The impact of the proposed segmentation algorithms is evaluated on the speaker diarization task using the VBx system \cite{landini_bayesian_2022}.
Results show that the proposed extensions offer competitive performance considering standard beamforming and SACC.
Finally, systems trained on multichannel data often suffer performance degradation in case of a mismatch in the array geometry between training and evaluation sets.
We propose a channel-number invariant loss to facilitate the model generalization under mismatched array configuration.
To the best of our knowledge, no similar training procedure has yet been applied to VAD and OSD.
The source code will be made available in a large-scale diarization toolkit to be published soon.

The paper is organized as follows. 
Section~\ref{sect:2_desc_tasks} formulates the joint VAD+OSD task. 
Section~\ref{sect:3_mch_feats} introduces the proposed algorithms. 
The systems and baselines are described in Section \ref{sect:baselines}, followed by the experimental protocol detailed in Section~\ref{4_sect_protocol}.  
Section \ref{5_sect_results_seg} evaluates the VAD and OSD results while speaker diarization results are described in Section \ref{6_sect_results_diar}.
The analysis of the spatial information used by complex systems is conducted in Section~\ref{7_sect_csacc_detail}.
Finally, the channel-number invariant loss is introduced and evaluated in Section~\ref{8_sect_invariance} before drawing conclusions and perspectives in Section~\ref{9_sect_conclusion}.

%% file: 02_speech_segmentation.tex
\section{Voice Activity and Overlapped Speech Detection}

\label{sect:2_desc_tasks}

This paper focuses on two speech segmentation sub-tasks, namely Voice Activity Detection (VAD) and Overlapped Speech Detection (OSD). 
The former aims to detect speech segments in the signal, while the latter aims to find segments in which at least two speakers are simultaneously active. 
VAD and OSD can be solved within the same framework, which is detailed in the following sections.

\begin{figure}[b!]
    \centering
    \includegraphics[width=0.8\linewidth]{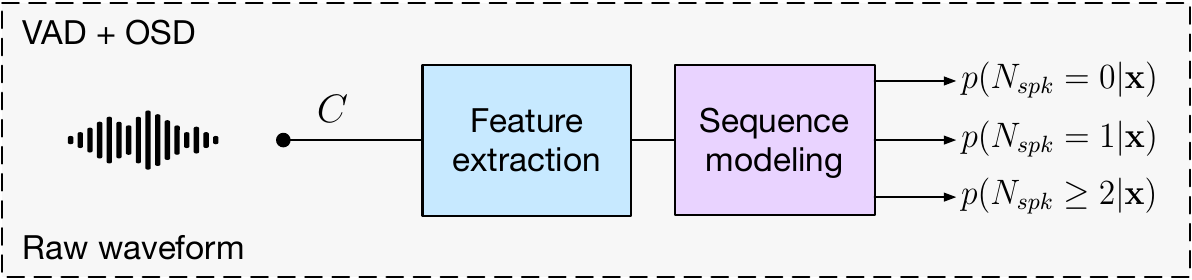}
    \caption{Simplified sequence-to-sequence model architecture for joint VAD+OSD. The feature extraction is different between the single channel $C=1$ and the multi-channel $C>1$ scenario. The sequence modeling step remains unchanged regardless the number of channels.}
    \label{fig:tasks}
\end{figure}

\subsection{General formulation}

Let $\mathbf{X}=[\mathbf{X}_0,\dots \mathbf{X}_t, \dots \mathbf{X}_{T-1}]\in \mathbb{R}^{F\times T}$ be a sequence of feature vectors where $F$ is the number of features, $T$ the number of time frames and $t$ the time frame index. 
This sequence is extracted from the raw audio signal $\mathbf{x}\in\mathbb{R}^{C\times N}$,  with $C$ being the number of channels (i.e. microphones) and $N$ the number of samples.
Feature extraction can be defined as a function $g:\mathbf{x}\rightarrow\mathbf{X}$, which maps the raw input signal to a sequence of feature vectors $\mathbf{X}\in \mathbb{R}^{F\times T}$.
The $g$ function might be handcrafted or trained end-to-end with the sequence-modeling model described below. 
As proposed in \cite{bredin_pyannoteaudio_2020,cornell_detecting_2020}, the sequence of features verifies $T<N$ to reduce the computational cost.
Note that the $g$ function depends on the number of channels in the input signal $\mathbf{x}$.
The feature extraction differs between the single-channel, $C=1$, and the multi-channel scenario, $C>1$. 

Let $\mathbf{y}=[\mathbf{y}_0,\dots\mathbf{y}_t,\dots\mathbf{y}_{T-1}]\in \mathbb{R}^{ T}$ be a sequence of aligned reference binary labels. 
VAD and OSD are solved by optimizing the parameters $\hat{\boldsymbol{\theta}}$ of a model $f : \mathbf{X}, \boldsymbol{\theta} \rightarrow \hat{\mathbf{y}}$ which maps the feature sequence to a sequence of predicted labels $\hat{\mathbf{y}}=[\hat{\mathbf{y}}_0,\dots\hat{\mathbf{y}}_t,\dots\hat{\mathbf{y}}_{T-1}]\in \mathbb{R}^{ T}$.

\subsection{Labelling procedure}

VAD and OSD are solved jointly following a 3-class classification problem following \cite{jung21_interspeech,Cornell2022}.
The neural model is trained to predict a sequence $\hat{\mathbf{y}}$ from the raw input signal $\mathbf{x}$ based on the manual annotations $\mathbf{y}$.

To solve VAD and OSD tasks, the frame-level annotation verifies $y_t\in\{0,1,2\}$.
$y_t=0$ refers to a non-speech frame ($N_{spk}=0$).
$y_t=1$ indicates that the $t$-th frame contains speech from a single active speaker ($N_{spk}=1$).
$y_t=2$ refers to a frame containing at least two simultaneously active speakers ($N_{spk}\geq 2$).

Since overlapping speech is a rare event, the classes are unbalanced \cite{lebourdais2022overlaps}.
The class balance can be improved by artificially generating additional overlapped data by combining single-speaker utterances from other datasets \cite{cornell_detecting_2020,kunevsova2019detection} or random segments of the training data at training time \cite{bullock_overlap-aware_2020}.
Here, we consider the second data augmentation strategy.
Figure~\ref{fig:tasks} summarizes the flowchart of the VAD+OSD model.  
The system predicts 3 frame-level pseudo probabilities, namely non-speech $p(N_{spk}=0|\mathbf{x})$, single active speaker $p(N_{spk}=1|\mathbf{x})$ and multiple active speakers $p(N_{spk}\geq 2|\mathbf{x})$.
VAD prediction is the combination of the $N_{spk}=1$ and the $N_{spk}\geq2$ outputs, while OSD prediction is determined from the $N_{spk}\geq2$ output only.




%% file: 03_mch_feats.tex
\section{Multi-channel feature extraction}
\label{sect:3_mch_feats}
\begin{figure}[htbp]
    \centering
    \includegraphics[width=0.3\textwidth]{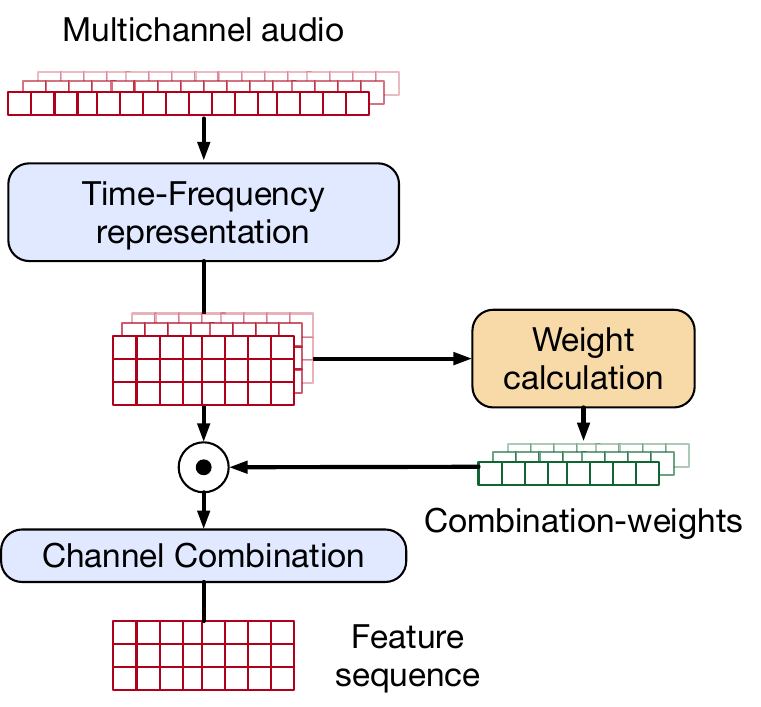}
    \caption{Channel combination procedure for multi-channel feature extraction. The time-frequency representation and the weight calculation algorithm depend on the algorithm considered (e.g. SACC).} 
    \label{fig:global_archi}
\end{figure}

In the meeting context, capturing distant speech signals offers practical benefits by preventing the participants from carrying individual microphones.
Distant speech signals however, tend to degrade the performance of automatic speech processing systems~\cite{maciejewski_characterizing_2018}.
We investigate several multi-channel front-ends to tackle distant VAD and OSD.
The proposed approaches consist of weighting and combining the incoming channels to generate an enhanced single-channel feature sequence.
All the implemented algorithms are inspired by the Self Attention Channel Combinator (SACC)~\cite{gong_self-attention_2021}. 
Figure~\ref{fig:global_archi} depicts the global approach for multi-channel feature extraction.
Specific weight estimation algorithms are presented in the following sections.

\subsection{Self Attention Channel Combinator}

The Self-Attention Channel Combinator (SACC)~\cite{gong_self-attention_2021} was developed in the context of multi-channel speech recognition as an alternative to neural network-based beamforming~\cite{li2016neural}.
This approach consists of estimating time-varying weights, referred to as \textit{combination weights}, to weight and to combine channels captured by a set of microphones. 
The weight estimation is performed in the Short-Time Fourier Transform (STFT) domain using a self-attention module~\cite{vaswani_attention_2017}. 
The model is thus supposed to focus attention on the microphones containing relevant information for the back-end task.

\begin{figure*}[hb]
    \centering
    \includegraphics[width=\linewidth]{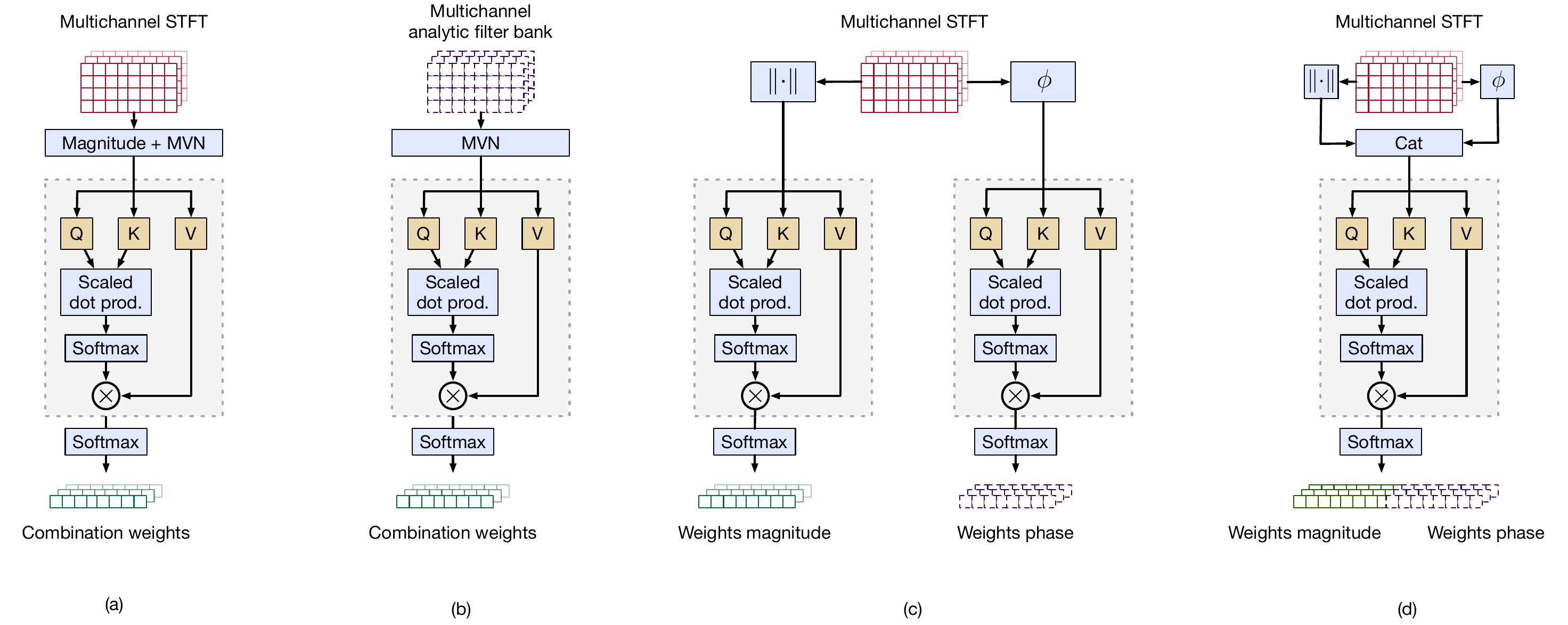}
    \caption{Weight calculation procedures, with (b) to (d) being the proposed methods. (a) STFT-based SACC~\cite{gong_self-attention_2021}, (b) SACC based on the analytic filter bank, (c) EcSACC, and (d) IcSACC. For complex systems (c) and (d), magnitude and phase ($\|\cdot\|$-$\phi$)  can be replaced by real and imaginary parts ($\mathfrak{R}$-$\mathfrak{I}$).}
    \label{fig:all_sacc}
\end{figure*}

Let $\mathbf{Y} \in \mathbb{C}^{C\times T\times K}$ be the multichannel STFT of the input signal $\mathbf{x}$, with $K$ being the number of frequency bins. 
The SACC algorithm computes combination-weights, $\mathbf{w}\in \mathbb{R}^{C\times T\times 1}$, from the magnitude of the multichannel STFT using self-attention \cite{vaswani_attention_2017}. 
Let $q$, $k$, and $v$ be three linear transformations, mapping the STFT log-magnitude $\log(\|\mathbf{Y}\|)$ to the query and the key $\mathbf{Q},\mathbf{K}\in\mathbb{R}^{C\times T\times D}$, and the value $\mathbf{V}\in\mathbb{R}^{C\times T\times 1}$. The combination weights are determined as follows:

\begin{equation}
    \mathbf{w}=\mathrm{softmax}\bigg(\mathrm{softmax}\Big(\frac{\mathbf{Q}\mathbf{K}^T}{\sqrt{D}}\Big)\mathbf{V}\bigg),
    \label{eq:comb_weight}
\end{equation}

\noindent with $D$ being the output space dimension of the query $\mathbf{Q}$ and the key $\mathbf{K}$. 
Note that the transpose operator $.^T$ is applied framewise. 
Each $C\times T$ matrix is transposed for each time frame.
The last $\mathrm{softmax}$ activation applies to the channel dimension and constrains the weights to be within the interval $[0,1]$. 
Mean and Variance Normalization (MVN) is applied to each frequency bin before feeding the self-attention module. 
MVN reduces the data variation range to make combination-weights learning easier.
The combined time-frequency representation $\mathbf{Y}_{att} \in \mathbb{R}^{1\times T\times K}$ is finally obtained as the weighted sum of the different channels:
\begin{equation}
    \mathbf{Y}_{att}=\sum_{c=1}^C\mathbf{w}\odot \|\mathbf{Y}\|,
    \label{eq:combination}
\end{equation}

\noindent with $\odot$ being the element-wise product. 
The combined STFT, $\mathbf{Y}_{att}$, is finally converted to the log-mel scale using $F$ mel scale filters to obtain the feature vector $\mathbf{X}  \in \mathbb{R}^{T\times F}$.
The process of the SACC algorithm is presented in Figure \ref{fig:all_sacc}(a). 
This model is referred to as SACC+$STFT$.

\subsection{Analytic SACC}

In the original SACC implementation~\cite{gong_self-attention_2021}, the weights are computed from the STFT magnitude. 
The phase information contained in the STFT is lost at the beginning of the process. 
This information might however be relevant for the weights estimation process since it informs on the time delay between channels.
Including the phase-related information may improve the weight estimation procedure.

To implicitly include phase information during weight estimation, we propose to replace the STFT with a learnable filter bank. 
The input signal is decomposed into several frequency bands with a time domain filter bank.
Filters are implemented as a set of mono-dimensional convolutional layers optimized in an end-to-end manner with the back-end task, i.e. VAD and OSD. 
Learning consistent filters from free convolutional layers may however be difficult when dealing with high-order filters. 
To counteract this limitation, learnable analytical filters have been proposed in the context of speech separation \cite{pariente2020filterbank}.

Let $\mathbf{h}(n) \in \mathbb{R}^{L}$ be the impulse response of a real filter with $n$ being the sample index and $L$ its length. 
The analytical version of this filter can be computed through the Hilbert transform $\mathcal{H}$ as follows:

\begin{equation}
    \mathbf{h}_{analytic}(n)=\mathbf{h}(n) + j\mathcal{H}\{\mathbf{h}(n)\},
\end{equation}

\noindent where $j=\sqrt{-1}$. 
In practice, the real and imaginary parts of $\mathbf{h}_{analytic}(n)$ are concatenated before computing the combination weights.

In the analytical version of the SACC model, the STFT is replaced by a set of analytical filters.
The filter bank is shared across channels to reduce the number of trainable parameters.
The output STFT-like representation is then forwarded through the self-attention module as depicted in figure \ref{fig:all_sacc}(b).
This model is referred to as SACC+$\mathcal{A}$.

\subsection{Complex SACC}

Another approach to take the phase information into account is to integrate the phase of the STFT in the combination-weight estimation procedure.
The combination weights thus belong to the complex space, i.e. $\mathbf{w}\in\mathbb{C}^{C\times T\times 1}$.
This approach was proposed in~\cite{mariotte22_interspeech} to learn combination weights that are closer to the conventional beamforming techniques.
Furthermore, this formulation allows us to better interpret the weights learned by the model as shown in section~\ref{7_sect_csacc_detail}.

This paper investigates two formulations to estimate the complex combination weights.
The first computes the magnitude and the phase of the weights separately using two distinct self-attention modules applied to the magnitude (Mag) and the phase ($\phi$) of the STFT respectively.
The second estimates the complex weights using a single self-attention module applied to the concatenation of the magnitude and the phase. 
These approaches are respectively referred to as Explicit complex SACC (EcSACC) and Implicit complex SACC (IcSACC).

\subsubsection{Explicit cSACC}


EcSACC consists of computing the magnitude and the phase of the combination weights separately. 
Each component of these weights is thus computed explicitly.
The system flowchart is presented in figure~\ref{fig:all_sacc}(c).
Let $\|\mathbf{Y}\|$ and $\angle \mathbf{Y}$ be the magnitude and the phase of the multi-channel STFT. 
The weights associated with each part of the STFT are computed as follows:

\begin{equation}
\begin{matrix}
\mathbf{w}_{Mag} = \mathrm{SA}_{Mag}\left(\|\mathbf{Y}\|\right), & \mathbf{w}_{\phi} = \mathrm{SA}_{\phi}\left(\angle \mathbf{Y}\right),\\
\end{matrix}
\label{eq:ecsacc_comb_weights}
\end{equation}
where $\mathrm{SA}_{Mag}$ and $ \mathrm{SA}_{\phi}$ represent the self-attention modules defined in equation \eqref{eq:comb_weight} applied to the magnitude and the phase respectively. 
The magnitude $\mathbf{w}_{Mag}$ and phase $\mathbf{w}_{\phi}$ of the weights are then applied to the associated parts of the multichannel STFT:

\begin{equation}
    \mathbf{Y}_{att} =\left(\mathbf{w}_{Mag}\odot \|\mathbf{Y}\|\right)e^{j(2\pi \mathbf{w}_{\phi} + \angle \mathbf{Y})}.
    \label{eq:EcSACC_combination}
\end{equation}
The phase weights verify $\mathbf{w}_{\phi}\in[0,1]$ with the $\mathrm{softmax}$ normalization in Eq.\eqref{eq:comb_weight}.
The $2\pi$ factor ensures that the phase correction covers the entire unit circle.
Note that there is no weight sharing between the two self-attention modules to enable the model to learn two distinct representations.
The same weight-estimation procedure can be applied to the real ($\mathfrak{R}$) and the imaginary ($\mathfrak{I}$) parts of the STFT.\\

\subsubsection{Implicit cSACC}

EcSACC processes the magnitude and the phase separately, which prevents the model from learning a relationship between the two representations.
Furthermore, it requires learning two self-attention modules which doubles the number of trainable parameters in the feature extraction.
The IcSACC formulation can learn cross-relationships between each part of the STFT while keeping the same number of trainable parameters as SACC. 
The IcSACC approach is presented in figure~\ref{fig:all_sacc}(d). 
It consists of a single self-attention module that estimates the combination weights from the concatenation of the magnitude and the phase of the STFT:

\begin{equation}
    \mathbf{w} = \mathrm{SA}\left(\mathrm{cat}(\|\mathbf{Y}\|,\angle \mathbf{Y})\right),
    \label{eq:IcSACC_weights_est}
\end{equation}

\noindent where $\mathrm{cat}(\cdot,\cdot)$ represents the concatenation operator. 
The two STFT parts are concatenated over the channel dimension. 
The resulting vector $\mathbf{w}$ contains the concatenation of the magnitude and the phase of the weights. 
The channel combination is then performed following equation~\eqref{eq:EcSACC_combination}. 


%% file: 04_protocol.tex
\section{Systems and baselines}
\label{sect:baselines}
This section describes the baseline systems along with the hyperparameters used for each architecture.

\subsection{Feature extraction configuration}

SACC-based models are compared to two single-channel baselines.
The first is trained and evaluated on the close talk data captured by AMI \textit{headset-mix} while the second is trained and evaluated on the first channel of \textit{Array~1} (single distant microphone).
The close-talk baseline features an upper-bound score, considering that distant performance tends to be degraded~\cite{maciejewski_characterizing_2018}. 
A multi-channel baseline, based on Minimum Variance Distortion-less (MVDR) beamforming, is also considered.
The following sections describe the feature extraction procedure for each system.
The length of the feature vector has been optimized separately for each feature extractor.
Thus, the number of features $F$ differs from one system to another.\\

\subsubsection{Single channel}

The Mel Frequency Cepstral Coefficients (MFCC) are used as single-channel features.
They have been extensively used for VAD and OSD in the literature \cite{bredin_pyannoteaudio_2020,bullock_overlap-aware_2020}.
In this study, 20 MFCC are extracted from the raw audio signal using a 25ms window with a 10ms shift.
The first $\Delta$ and second $\Delta\Delta$ derivatives are also computed.
The first coefficient is removed resulting in a feature vector composed of $F=59$ coefficients.

\subsubsection{MVDR Beamformer}
To compare SACC-based approaches to standard beamforming algorithms, the Minimum Variance Distortion-less (MVDR) beamformer proposed in~\cite{sounden_optimal_2010} is also implemented.
This method requires estimating the noise covariance matrix.
Since we are dealing with realistic speech conditions, the clean speech of the source is unknown.
We use the Coherent-to-Diffuse Ratio (CDR) under diffuse noise and unknown DOA conditions --see \cite{schwarz_coherent_diffuse_2015} Eq. (25)-- to estimate the time-frequency masks.
The noise covariance matrix is then estimated from the masked signal.
Beamformer weight estimation and channel combination are performed in the STFT domain.
Similarly to STFT-based SACC, the resulting spectrogram is converted to the Mel scale using $F=64$ filters.

\subsubsection{Self-Attention Channel Combinators}
Several multi-channel front-ends are considered in this study.
The STFT-based features -- SACC, EcSACC, and IcSACC -- are extracted using 25ms windows with a 10ms shift.
For each SACC variant, the hidden size of the self-attention module is set to $D=256$.
The output spectrograms of SACC+$STFT$, EcSACC, and IcSACC are converted to the Mel scale with $F=64$ filters.
In the SACC+$\mathcal{A}$ model, features are extracted with 32 analytic filters.
The concatenation of the real and imaginary parts of the filter's output results in a vector of $F=64$ features.
The kernel of the convolution layer is designed to fit the STFT parameters with a window size of 25ms and a shift of 10ms.

\subsection{Sequence modeling architecture}

The sequence of features extracted from the raw audio signal is processed by a sequence modeling network to predict VAD+OSD labels.
To better evaluate the impact of each feature extractor on each task performance, two sequence modeling architectures are considered.
The first is based on Bidirectional Long Short-Term Memory (BLSTM) while the second is a Temporal Convolutional Network (TCN). 
Architecture details are presented in the following sub-sections.

\subsubsection{BLSTM}

The former sequence modeling architecture composed of two stacked BLSTM layers is composed of 256 cells similar to \cite{bullock_overlap-aware_2020,bredin_pyannoteaudio_2020}. 
The output sequence is post-processed using a three-layer feed-forward network (FFN) with output sizes $L_1=128$, $L_2=128$, and $L_3=2$ respectively. 
The last layer outputs the pseudo-probability of each class.
FFN layers are followed by a $\mathrm{tanh}$ activation function except for the last one. 
A $\mathrm{softmax}$ activation is applied to the output logits to compute the classification scores.

\subsubsection{TCN}

The Temporal Convolutional Network (TCN) has been proposed as an alternative to BLSTM for VAD and OSD \cite{cornell_detecting_2020}. 
This architecture is composed of causal convolutional layers with residual connections. 
The TCN uses dilated convolutions to benefit from a large temporal context.
Our system features the same TCN architecture as~\cite{cornell_detecting_2020}.
A layer normalization step is first applied, followed by a 1D-convolution bottleneck layer which projects the sequence of features from $F$ dimensions to 64 dimensions.
The bottleneck output is then processed by 5 1-D convolution layers of 128 hidden channels, with exponentially increasing dilatation.
This block is repeated 3 times. 
A residual connexion is added at the output of each block.
Finally, a 1-D convolutional layer projects the hidden sequence to the classification space (3 dimensions). 

\section{Experimental protocol}
\label{4_sect_protocol}

Our experiments aim to evaluate how each multi-channel front end impacts VAD and OSD performance.
This section introduces the dataset and the evaluation protocol.

\subsection{Dataset}

Experiments are conducted on the AMI meeting corpus~\cite{Mccowan05theami}. 
This dataset features about 100 hours of meetings recorded under realistic conditions. 
Audio signals have been captured using different devices. 
In this work, the signals recorded by headset microphones, referred to as \textit{headset-mix}, are considered as a close-talk reference. 
Multichannel signals acquired by the 8-microphone array placed in the center of the table referred to as \textit{Array 1}, are used as distant speech material. 
This device consists of a uniform circular array (UCA), with $r=0.1$m radius.
Distant audio signals are also captured by a 4-microphone array, referred to as \textit{Array 2}.
Both 8- and 4-microphone arrays are considered to evaluate channel-number invariant systems (section~\ref{8_sect_invariance}). 
This last device is either circular or linear depending on the meeting session.

Framewise VAD+OSD labels are extracted from the speaker diarization annotation provided in the AMI dataset. 
Overlapping speech labels are obtained where two annotated segments overlap.
The data is divided into training, development, and test sets based on the protocol proposed in~\cite{landini_bayesian_2022}, which guarantees different speakers in each set. 
10\% of the training set is reserved for validation to determine the best-performing system.
To reduce the computational cost, the label sequence is sampled at 100Hz, while the audio signal has a 16kHz sampling rate.

\subsection{Training and evaluation procedure}

VAD+OSD systems are trained on 2000 batches of 64 segments per epoch. 
Training segment duration is set to 2s.
These segments are randomly sampled in the AMI training set.
The training stops after five epochs if there is no improvement in the target validation metric (F1-score).
Since VAD+OSD is formulated as a 3-class classification problem, cross-entropy $\mathcal{L}_{CE}$ is used as a training objective. 
A channel-number invariant loss $\mathcal{L}_{inv}$ can also be added as detailed in section~\ref{8_sect_invariance}.
Model optimization is performed using the ADAM optimizer~\cite{kingma2014adam} with a $10^{-3}$ learning rate.
The validation is conducted on batches of 64 segments from the validation set.
The best model is selected as the one obtaining the higher F1-score on the validation set.

The evaluation is conducted on the development (Dev) and evaluation (Eval) sets of the AMI meeting corpus. 
The inference is performed on a 2s sliding window with a 0.5s shift.  
Predictions are averaged on the overlapped part of the sliding windows.
For each frame, the predicted class is determined by finding the $\mathrm{argmax}$ among the pseudo-probabilities.
The detected segments are saved in the Rich Transcription Time Marked (RTTM) format which allows the evaluation of VAD and OSD with the reference segmentation.
VAD is evaluated using the false alarm rate (FA), the miss detection rate (Miss), and the sum of these two metrics: the Segmentation Error Rate (SER). 
OSD is evaluated using precision, recall, and F1-score. 


%% file: 05_seg_results.tex
\section{Segmentation results}
\label{5_sect_results_seg}

This section presents the VAD and OSD performance of both BLSTM and TCN architectures with each front end.
Results are presented on the development and evaluation sets of the AMI corpus.

\subsection{BLSTM results}

Table \ref{tab:vad_osd_blstm} presents the VAD and OSD performance of the 3-class BLSTM system.
The close-talk model reaches an OSD F1-score of 70.1\% and 69.2\% on Dev and Eval sets, respectively.
This model achieves a VAD SER of 6.66\% and 5.94\% on these two subsets.
This score is the upper bound baseline with the BLSTM sequence modeling.
Single-channel distant speech degrades the performance, with 64.0\% and 62.5\% OSD F1-score.
VAD performance is also lower, e.g. 7.06\% on the evaluation set.
This shows how distant speech conditions degrade the segmentation.
The SDM acts as a lower-bound reference.

Considering multichannel processing such as MVDR drastically improves distant OSD with an F1-score of 66.9\% on the Eval.
VAD also improves with a 6.55\% evaluation SER. 
The original SACC model shows better performance, with a 67.4\% OSD F1-score and a 6.40\% VAD SER on the evaluation subset.
It offers the best VAD and OSD performance on the evaluation set with the BLSTM architecture.

Among the proposed models, the SACC+$\mathcal{A}$ degrades the detection performance on both OSD (64.5\% F1-score Eval) and VAD (6.71\% SER Eval) compared to SACC.
The performance however remains higher than the SDM scenario.
IcSACC shows similar OSD performance as $\mathcal{A}$ on the development set (66.7\%) but improves on the evaluation data with 66.0\%.
The VAD performance of this system is slightly better than the SDM with 6.73\% on the evaluation set.
The EcSACC formulation shows similar OSD performance as MVDR, with 68.0\% on Dev and 66.5\% on Eval.
The VAD scores remain degraded with 6.80\% on the evaluation set.

\begin{table*}[ht]
    \centering
    \caption{VAD and OSD performance of the 3-class BLSTM segmentation model. The results are obtained on the AMI development and evaluation sets.  Bold values indicate the best systems, i.e. within the confidence interval of the best system estimated at the file level. Pr: precision, Re: recall.}
    \resizebox{0.9\textwidth}{!}{
    \begin{tabular}{lcccccccccccc}
    \toprule
    \multirow{3}{*}{\textbf{BLSTM}} & \multicolumn{6}{c}{\textbf{OSD}} & \multicolumn{6}{c}{\textbf{VAD}} \\
    \cmidrule(lr){2-7}
    \cmidrule(lr){8-13}
    & \multicolumn{2}{c}{Pr.$_{\%\uparrow}$} & \multicolumn{2}{c}{Re.$_{\%\uparrow}$} & \multicolumn{2}{c}{F1$_{\%\uparrow}$} & \multicolumn{2}{c}{FA$_{\%\downarrow}$} & \multicolumn{2}{c}{Miss$_{\%\downarrow}$} & \multicolumn{2}{c}{SER$_{\%\downarrow}$}\\
    & Dev & Eval & Dev & Eval & Dev & Eval & Dev & Eval & Dev & Eval & Dev & Eval \\
    \midrule
    Close talk & \textit{64.8} & \textit{72.8} & \textit{76.4} & \textit{66.0} & \textit{70.1} & \textit{69.2}  & \textit{5.35} & \textit{3.85} & \textit{1.32} & \textit{1.79} & \textit{6.66} & \textit{5.64}\\
    SDM  & 59.5 & 63.0 & 69.3 & 61.9 & 64.0 & 62.5 & 4.13 & 3.70 & 2.54 & 3.39 & 6.68 & 7.06\\
    MVDR & 64.0 & 70.7 & 73.7 & 63.5 & 68.5 & 66.9 & 4.04 & 3.91 & 1.87 & 2.64 & \textbf{5.91} & \textbf{6.55}\\
    SACC+$STFT$ & 69.2 & 71.9 & 71.1 & 63.4 & \textbf{70.2} & \textbf{67.4} & 4.52 & 4.13 & 1.53 & 2.27 & \textbf{6.05} & \textbf{6.40}\\
    SACC+$\mathcal{A}$ & 66.6 & 68.2& 67.6 & 61.2 & 67.1 & 64.5 & 5.22 & 4.65 & 1.58 & 2.06 & 6.80 & 6.71\\
    IcSACC Mag-$\phi$   & 61.4 & 67.9 & 72.9 & 64.2 & 66.7 & 66.0 & 4.44 & 3.68 & 1.99 & 3.04 & 6.43 & 6.73\\
    EcSACC Mag-$\phi$   & 64.8 & 71.6 & 71.6 & 62.1 & 68.0 & 66.5 & 4.39 & 3.58 & 1.96 & 3.23 & 6.35 & 6.80\\
    
    \bottomrule
    \end{tabular}
    }
    \label{tab:vad_osd_blstm}
\end{table*}

\subsection{TCN results}
Table \ref{tab:vad_osd_tcn} shows each front-end performance on VAD and OSD by considering the TCN architecture.
The close-talk model offers a 74.0\% F1-score on the evaluation set.
Considering a TCN instead of a BLSTM improves OSD by an absolute +4.8\% on this subset.
VAD SER also improves from 6.66\% to 5.42\% on the development set.
The SDM system degrades OSD (64.9\% Eval F1-score) and VAD (6.81\% Eval SER) because of distant recording conditions.

MVDR offers remarkable OSD performance with 72.3\% and 69.6\% F1-score on development and evaluation sets respectively.
This model also shows the best distant VAD performance (6.32\% Eval SER).
The SACC model offers similar OSD performance with 71.8\% and 68.5\% F1-score on the two subsets but highlights a lack of robustness on VAD with 6.73\% on the evaluation set for the SER.

Like the BLSTM, SACC+$\mathcal{A}$ shows limited OSD capacities with 64.6\% F1-score on the evaluation set.
However, the VAD improves concerning the SDM and the original SACC (6.58\% Eval SER).
IcSACC gets OSD performance close to MVDR and SACC on the Dev set (71.5\%). 
This model shows limited generalization capacities, with a 67.6\% F1-score on the evaluation set.
This degradation is mainly due to a drop in the recall.
The EcSACC model shows similar OSD performance as SACC on both OSD (68.4\% Eval F1-sore) and VAD (6.79\% SER Eval).

The original SACC and its complex extensions reach competitive performance with the MVDR beamformer.
The use of self-attentive methods remains justified since they do not require additional computations such as covariance matrix estimation.

While these two models do not significantly improve OSD compared to MVDR or SACC, they achieve similar performance, mainly on OSD. 
EcSACC and IcSACC improve speaker diarization under distant speech conditions and show strong explainability features, as shown in the next sections.

\subsection{Performance comparison}

This section compares the VAD and OSD performance of the three best TCN-based systems --SACC+STFT, IcSACC, and EcSACC-- with the best model from Cornell et al. \cite{Cornell2022}.
It consists of a VAD+OSD TCN model fed with log-scale Mel spectrogram and Cosine and Sine encoded Interaural Phase Difference (CSIPD) features.
The systems are compared in terms of Average Precision (AP) metric, as used in \cite{Cornell2022}.
Table \ref{tab:osd_cmp} presents the results obtained on the evaluation set from the AMI corpus.
It shows that our systems improve the VAD performance, as IcSACC and SACC+$STFT$ reach 99.3\% AP where \cite{Cornell2022} reaches 98.7\%.
Our systems also improve the OSD performance.
Specifically, SACC+$STFT$ shows the best performance with 71.4\% whereas the system from \cite{Cornell2022} reaches 60.4\%.

Note that the AMI protocol has been ambiguous in the context of speaker diarization, mostly due to the several versions proposed in the original paper \cite{Mccowan05theami}. 
The work from Landini et al. \cite{landini_bayesian_2022} proposes a new protocol for speaker diarization which meets consensus in the community.
As stated in section \ref{4_sect_protocol}, we used this protocol in this work.
However, Cornell et al. \cite{Cornell2022} use a different protocol.
Thus, the results are given for information only, and are not directly comparable with ours.

\begin{table}[b]
    \centering
    \caption{VAD and OSD Average Precision (AP$_{\%\uparrow}$) score on the AMI Eval subset of the three best SACC-based models compared with the best system from \cite{Cornell2022}. $^*$ This score is obtained with a different protocol for AMI partition. The comparison has to be taken with care.}
    \begin{tabular}{ccccc}         
         \toprule
         \textbf{Task} & \textbf{IcSACC} & \textbf{EcSACC} & \textbf{SACC+\textit{STFT}} & \textbf{Cornell et al. \cite{Cornell2022}} \\
         \midrule
         \textbf{VAD} & 99.3 & 99.2 & 99.3 & 98.7$^*$\\
         \textbf{OSD} & 70.6 & 70.9 & 71.4 & 60.4$^*$ \\
         \bottomrule         
    \end{tabular}    
    \label{tab:osd_cmp}
\end{table}

%% file: 06_diarization.tex
\section{Impact of the segmentation on speaker diarization}
\label{6_sect_results_diar}

\subsection{Protocol}
\label{diar_protocol}
The impact of each VAD+OSD model is evaluated on the final back-end task, i.e. speaker diarization.
A first diarization is obtained using the VBx system \cite{landini_bayesian_2022}.
This system uses a ResNet101 x-vector extractor followed by a VB-HMM clustering algorithm.
The VAD segments, predicted by our systems, are used as an initial segmentation.
X-vector clustering is initialized with Hierarchical Agglomerative Clustering (HAC) before performing VB clustering.

A second diarization is then obtained by assigning overlapping speech segments.
These segments are predicted by our OSD system.
The assignation is performed using the heuristic approach from \cite{otterson2007efficient}.
VBx speaker diarization and overlap assignment are performed using the online code available\footnote{\url{https://github.com/BUTSpeechFIT/VBx/tree/v1.1_VoxConverse2020} accessed on October 21st, 2023}.

The VAD and OSD segmentations are obtained with the TCN-based systems.
The diarization is evaluated with the Diarization Error Rate (DER) and the Jacquard Error Rate (JER) following the NIST evaluation protocols \cite{park2022review}.
The DER is computed both with a 25ms collar ($\delta_c=25$) and without ($\delta_c=0$).

\subsection{Impact of the VAD+OSD on speaker diarization}

Speaker diarization results are presented in table \ref{tab:chp3:diar_perf}.
The Oracle performance is obtained using the reference VAD and OSD labels. 
It represents the best score that can be expected with our diarization system, i.e., with the same segmentation as the annotation.
For each system, the speaker diarization results are presented with VAD only (VAD) and overlapping speech assignment (VAD w/ OSD).
Unless otherwise specified, we present DER performance in the $\delta_c=0$ evaluation setup, and the error between scores is relative.

The Oracle results show that the minimum DER that can be expected is 22.47\% with VAD only and 14.32\% by considering OSD.
Assigning the overlapping speech segments improves the diarization by +36.3\%, demonstrating that OSD is required for robust speaker diarization.

Among VAD+OSD systems, the SDM delivers the worst diarization performance, with 27.45\% DER with VAD only and 25.01\% with overlap assignments.
The performance gain due to OSD is reduced compared to Oracle with +8.9\% only.
The OSD errors have a strong impact on the speaker diarization performance.

Channel-combination front-ends have been shown to improve both VAD and OSD and improve speaker diarization.
SACC+$\mathcal{A}$ offers the highest DER among these systems with 23.97\%.
MVDR shows a 26.80\% and a 23.71\% DER with and without overlap assignment.
It reaches one of the best JER (30.87\%), highlighting a better intra-speaker segmentation.
SACC+$STFT$ achieves the lower DER with no collar in both scenarios (VAD: 26.06\%, VAD w/ OSD: 23.09\%).

Among the complex extensions, IcSACC achieves the best DER with a 25ms collar (15.87\%).
EcSACC offers the best diarization performance, as it reaches a similar JER as MVDR (30.90\%) and a DER close to SACC+$STFT$ (23.30\%).
This model exhibits a +11.6\% improvement between the VAD and the VAD w/ OSD scenarios.

The results obtained with SACC+$STFT$ and EcSACC also outperform both models from \cite{raj2022gpu}.
In this paper, the authors report a DER of 25.05\% with a VBx-based system and 23.69\% with a spectral clustering-based method.
For instance, the EcSACC system improves diarization of +7.0\% and +1.7\% respectively.

The experiments conducted on speaker diarization show that the performance gain in VAD+OSD leads to a gain in terms of DER and JER.
Thus, using channel combination approaches for VAD+OSD is also beneficial for the final task.
Complex models such as EcSACC and IcSACC achieve among the best diarization performance.

While showing competitive performance with MVDR and the original SACC, complex models offer new perspectives in terms of explainability.
The analysis of this system is presented in the next section.

\begin{table*}[ht]
    \centering
    \caption{VAD and OSD performance of the 3-class TCN segmentation model. The results are obtained on the AMI development and evaluation sets. Bold values indicate the best systems, i.e. within the confidence interval of the best system estimated at the file level. Pr: precision, Re: recall.}
    \resizebox{0.9\textwidth}{!}{
    \begin{tabular}{lcccccccccccc}
    \toprule
    \multirow{3}{*}{\textbf{TCN}} & \multicolumn{6}{c}{\textbf{OSD}} & \multicolumn{6}{c}{\textbf{VAD}} \\
    \cmidrule(lr){2-7}
    \cmidrule(lr){8-13}
    & \multicolumn{2}{c}{Pr.$_{\%\uparrow}$} & \multicolumn{2}{c}{Re.$_{\%\uparrow}$} & \multicolumn{2}{c}{F1$_{\%\uparrow}$} & \multicolumn{2}{c}{FA$_{\%\downarrow}$} & \multicolumn{2}{c}{Miss$_{\%\downarrow}$} & \multicolumn{2}{c}{SER$_{\%\downarrow}$}\\
    & Dev & Eval & Dev & Eval & Dev & Eval & Dev & Eval & Dev & Eval & Dev & Eval \\
    \midrule
    Close talk & \textit{69.9} & \textit{76.9} & \textit{78.8} & \textit{71.3} & \textit{73.8} & \textit{74.0} & \textit{3.23} & \textit{2.58} & \textit{2.18} & \textit{2.82} & \textit{5.42} & \textit{5.41} \\
    SDM  & 70.6 & 72.7 & 67.6 & 58.7 & 69.0 & 64.9 & 6.08 & 5.27 & 1.14 & 1.54 & 7.22 & 6.81 \\
    MVDR  & 72.5 & 77.0 & 72.3 & 63.6 & \textbf{72.3} & \textbf{69.6} & 3.69 & 3.21 & 2.25 & 3.12 & \textbf{5.94} & \textbf{6.32} \\
    SACC+$STFT$ & 72.2 & 74.8 & 71.5 & 63.2 & \textbf{71.8} & \textbf{68.5} & 2.89 & 2.43 & 3.04 & 4.31 & \textbf{5.94} & 6.73 \\
    SACC+$\mathcal{A}$ & 71.0 & 73.2 & 67.5 & 57.8 & 69.2 & 64.6 & 3.98 & 3.64 & 2.16 & 2.93 & \textbf{6.14} & \textbf{6.58} \\
    IcSACC Mag-$\phi$   & 71.6 & 75.6 & 71.5 & 61.1 & \textbf{71.5} &  67.6 & 4.10 & 3.36 & 2.13 & 3.41 & 6.24 & 6.77\\
    EcSACC Mag-$\phi$   & 73.0 & 76.9 & 70.3 & 61.7 & \textbf{71.6} & \textbf{68.4} & 3.50 & 2.85 & 2.60 & 3.94 & \textbf{6.10} & 6.79\\
    
    \bottomrule
    \end{tabular}
    }
    \label{tab:vad_osd_tcn}
\end{table*}

\begin{table}[b!]
    \centering
    \caption{Diarization performance of the VBx system with each distant segmentation model. Results obtained on the AMI evaluation set. $\delta_c$: duration of the collar applied during the evaluation in milliseconds.}
    \begin{tabular}{llccc}
        \toprule
        \multirow{2}{*}{\textbf{Segmentation}} & \multirow{2}{*}{\textbf{System}} & \multicolumn{2}{c}{\textbf{DER}$_{\%\downarrow}$} &  \multirow{2}{*}{\textbf{JER}$_{\%\downarrow}$} \\
        & & $\delta_c=25$ & $\delta_c=0$ & \\ 
        \midrule
        \multirow{2}{*}{Raj et al. \cite{raj2022gpu}} & VBx w/ OSD & - & 25.05 & - \\
        & Spectral w/ OSD & - & 23.69 & - \\
        \midrule
        \multirow{2}{*}{Oracle} & VAD     & \textit{15.63} & \textit{22.47} & \textit{30.52}\\
                                & VAD w/ OSD & \textit{10.12} & \textit{14.32} & \textit{25.78}\\
        \midrule
        \multirow{2}{*}{SDM}    & VAD     & 19.38 & 27.45 & 33.82\\
                                & VAD w/ OSD & 17.67 & 25.01 & 32.20\\
        \midrule
        \multirow{2}{*}{MVDR}    & VAD     & 18.19 & 26.80 & 32.77\\
                                 & VAD w/ OSD & 16.00 & 23.71 & \textbf{30.87}\\
        \midrule
        \multirow{2}{*}{SACC+$STFT$}    & VAD     &  18.21 & 26.06 & 32.83\\
                                        & VAD w/ OSD & 16.26 & \textbf{23.09} & 31.13\\
         \midrule
         \multirow{2}{*}{SACC+$\mathcal{A}$}   & VAD     & 17.92 & 26.51 & 33.02\\
                                                    & VAD w/ OSD & 16.29 & 23.97 & 31.60\\
         \midrule 
         \multirow{2}{*}{EcSACC} & VAD     & 18.52 & 26.35 & 32.54\\
                                 & VAD w/ OSD & 16.40 & \textbf{23.30} & \textbf{30.90}\\
         \midrule
         \multirow{2}{*}{IcSACC} & VAD     & 17.85 & 26.58 & 33.07\\
                                 & VAD w/ OSD & \textbf{15.87} & 23.73 & 31.40\\
         \bottomrule
    \end{tabular}
    \label{tab:chp3:diar_perf}
\end{table}

%% file: 07_cSACC.tex
\section{Complex weights analysis for explainability}
\label{7_sect_csacc_detail}

In Section~\ref{sect:3_mch_feats}, we introduced two extensions of the Self-Attention Channel Combinator to compute complex weights from the STFT.
This section proposes to apply a standard spatial filter analysis tool -- beampattern -- to show the explainability capacities of such systems.

\subsection{Beampattern as an analysis tool}

Similarly to beamforming, the complex combination weights allow the system to steer in a given angular direction.
Each complex SACC approach can be considered as a spatial filter.
The spatial response of such a filter can be visualized using the beampattern~\cite{benesty_design_2015}.
This representation informs on the system response to a plane wave impinging in a $\theta$ angular direction, given an array geometry.
Note that this representation can only be computed on complex combination weights.
Visualizing the magnitude of the beampattern for a set of combination weights indicates the spatial direction in which the self-attention module is focused. 
Considering a UCA and a set of learned combination-weights $\mathbf{w}_t$, the narrow-band beampattern can be computed as follows:

\begin{equation}
    \mathcal{B}_t[\mathbf{w}_t,\Bar{\omega},\theta]=\sum_{c=1}^C w_{c,t}\mathrm{e}^{j\Bar{\omega}\cos(\theta-\psi_c)},
    \label{eq:beampattern}
\end{equation}

\noindent where $t$ is the frame index, $c$ the microphone index, $\psi_c$ the angle of the $c$-th microphone in the array frame of reference, and $\Bar{\omega}=2\pi rf/v_{s}$ with $f$ being the frequency of the impinging plane wave, $r$ the radius of UCA and $v_{s}$ the speed of sound.


The broadband beampattern can then be computed for a given set of frequencies $\mathbf{f}=\{f_0,f_1,\cdots f_{sup}\}$, where $f_{sup}< \frac{C.v_{s}}{4\pi r}$ is the maximum frequency to avoid spatial aliasing~\cite{benesty_design_2015}.
The broadband beampattern is computed as follows:
\begin{equation}
    \boldsymbol{\mathcal{B}}_t[\mathbf{w}_t,\theta] = \Big[\mathcal{B}_t[\mathbf{w}_t,\Bar{\omega}_0,\theta], \cdots, \mathcal{B}_t[\mathbf{w}_t,\Bar{\omega}_{sup},\theta] \Big].
    \label{eq:broadband_bp}
\end{equation}

The EcSACC and IcSACC combination weights are time-variant. 
The beampattern also varies as a function of time.
The time average beampattern can be computed to visualize the average steering directions of the system on a given utterance:
\begin{equation}
    \bar{\mathcal{B}}[\mathbf{w}_t,\Bar{\omega},\theta] = \frac{1}{T} \sum_{t=0}^{T-1} \mathcal{B}_t[\mathbf{w}_t,\Bar{\omega},\theta],
    \label{eq:avg_p}
\end{equation}

\noindent with $T$ being the total number of time frames.
The following sections demonstrate the visualization capacities provided by the beampattern for the combination-weights analysis.

\subsection{Time averaged beampattern vs. acoustic energy map}

\begin{figure*}[t]
\begin{subfigure}{0.24\linewidth}
  \centering
  \includegraphics[trim=1cm 0 1cm 0,clip,width=\linewidth]{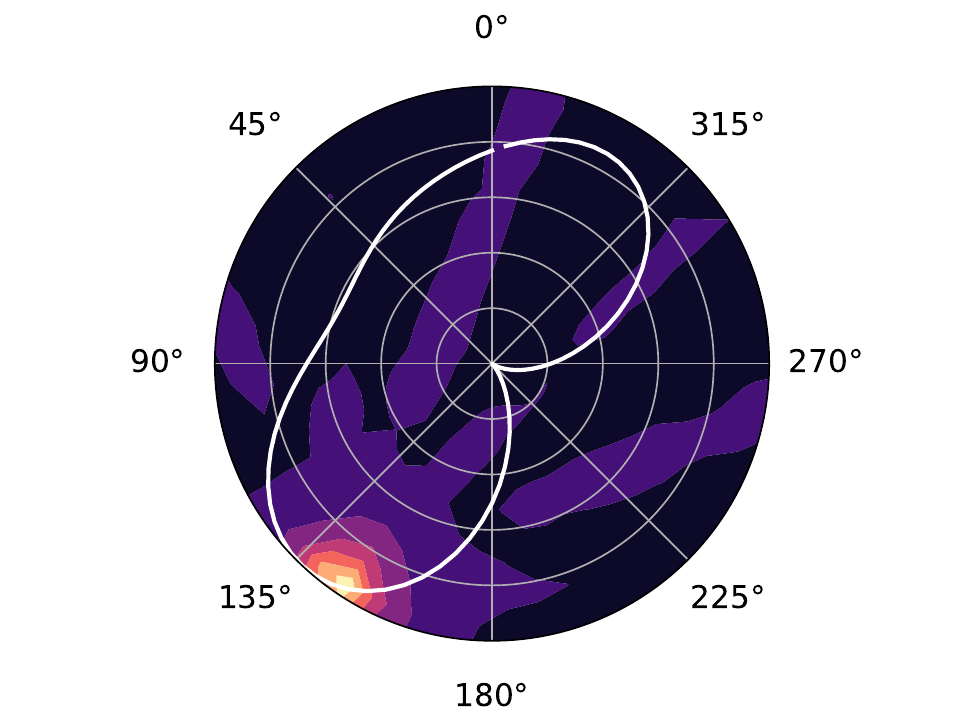} 
  \caption{Segment 1}
\end{subfigure}
\hfill
\begin{subfigure}{.24\linewidth}
  \centering
  \includegraphics[trim=1cm 0 1cm 0,clip,width=\linewidth]{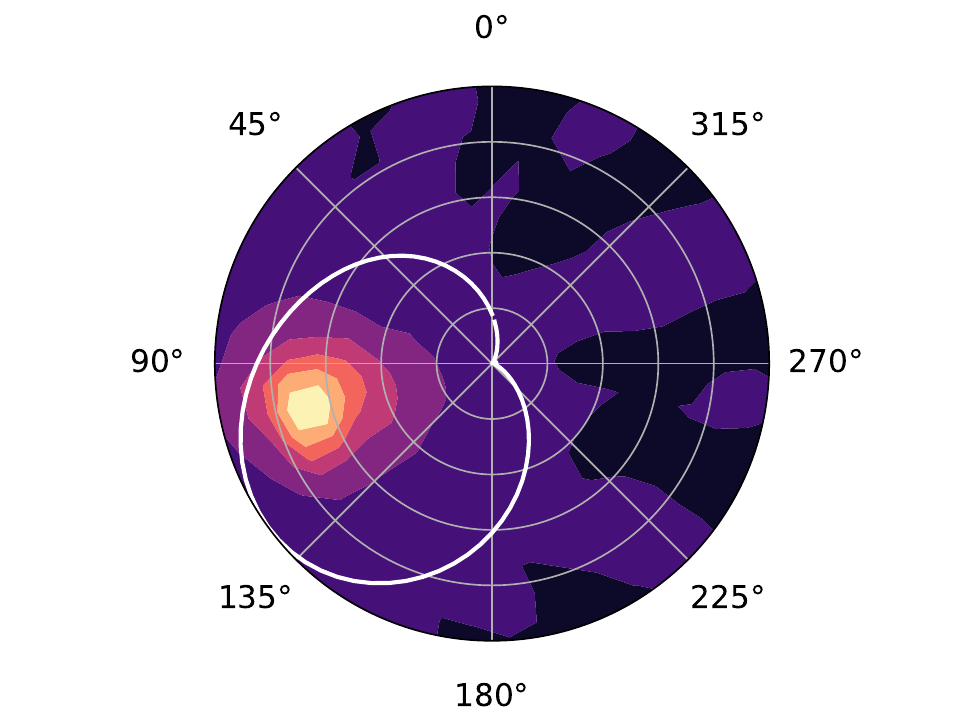}  
  \caption{Segment 2}
\end{subfigure}
\hfill
\begin{subfigure}{.24\linewidth}
  \centering
  \includegraphics[trim=1cm 0 1cm 0,clip,width=\linewidth]{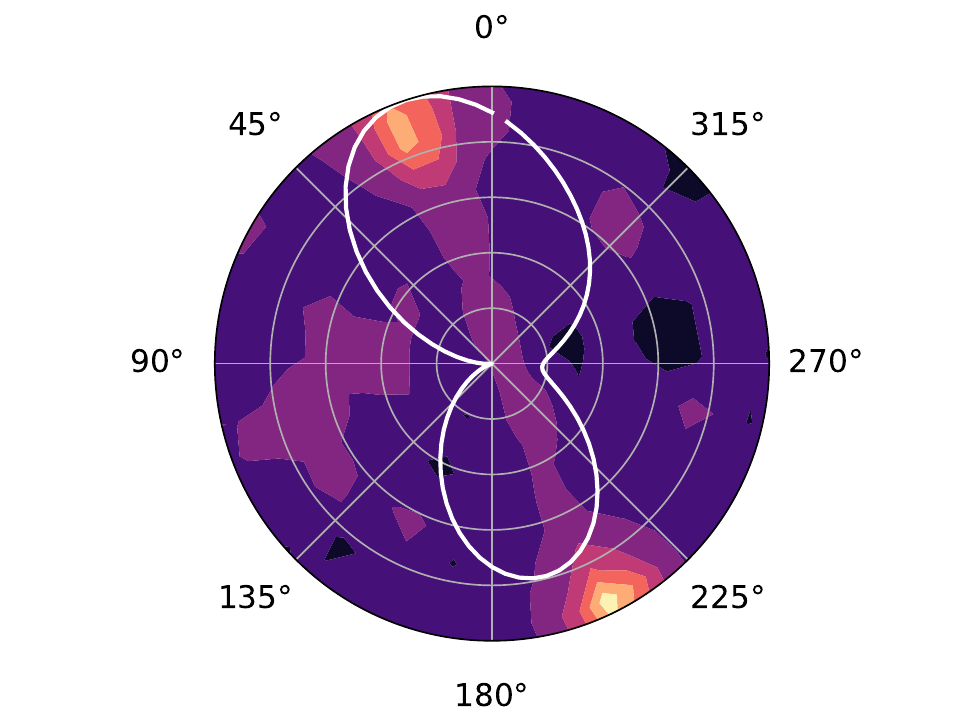} 
  \caption{Segment 3}
\end{subfigure}
\hfill
\begin{subfigure}{.24\linewidth}
  \centering
  \includegraphics[trim=1cm 0 1cm 0,clip,width=\linewidth]{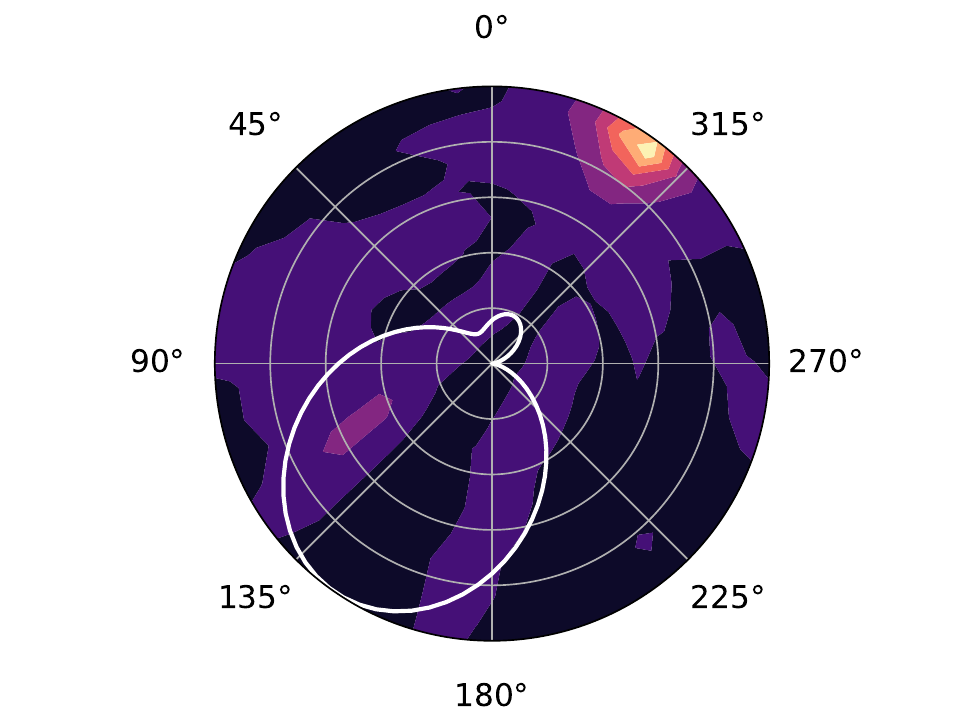}   
  \caption{Segment 4}
\end{subfigure}

\begin{subfigure}{0.24\linewidth}
  \centering
  \includegraphics[trim=1cm 0 1cm 0,clip,width=\linewidth]{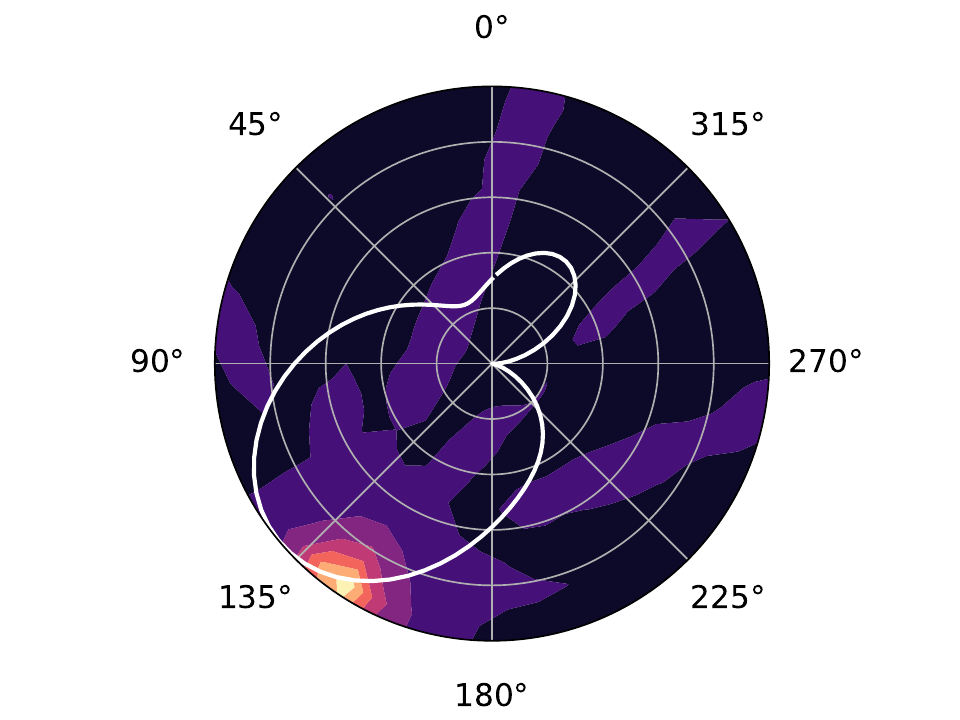} 
  \caption{Segment 1}
\end{subfigure}
\hfill
\begin{subfigure}{.24\linewidth}
  \centering
  \includegraphics[trim=1cm 0 1cm 0,clip,width=\linewidth]{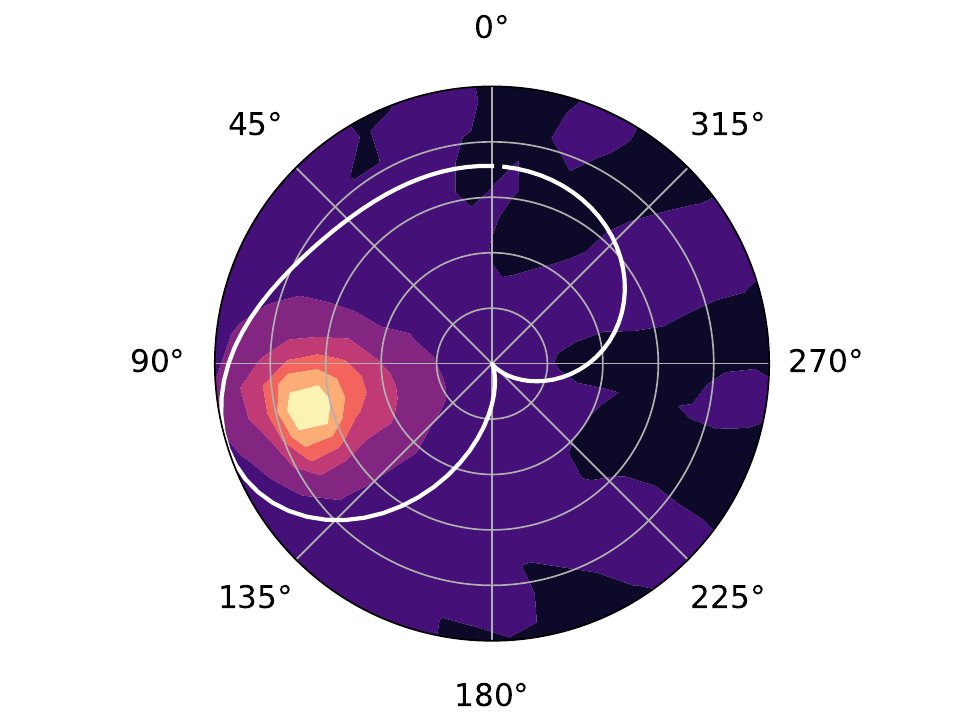}  
  \caption{Segment 2}
\end{subfigure}
\hfill
\begin{subfigure}{.24\linewidth}
  \centering
  \includegraphics[trim=1cm 0 1cm 0,clip,width=\linewidth]{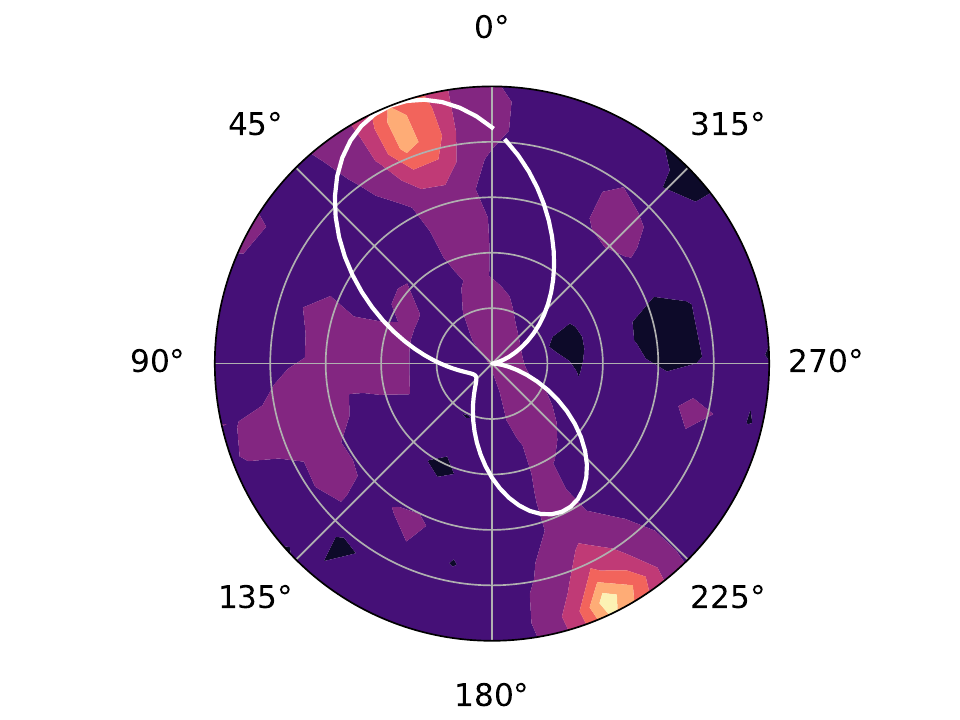} 
  \caption{Segment 3}
\end{subfigure}
\hfill
\begin{subfigure}{.24\linewidth}
  \centering
  \includegraphics[trim=1cm 0 1cm 0,clip,width=\linewidth]{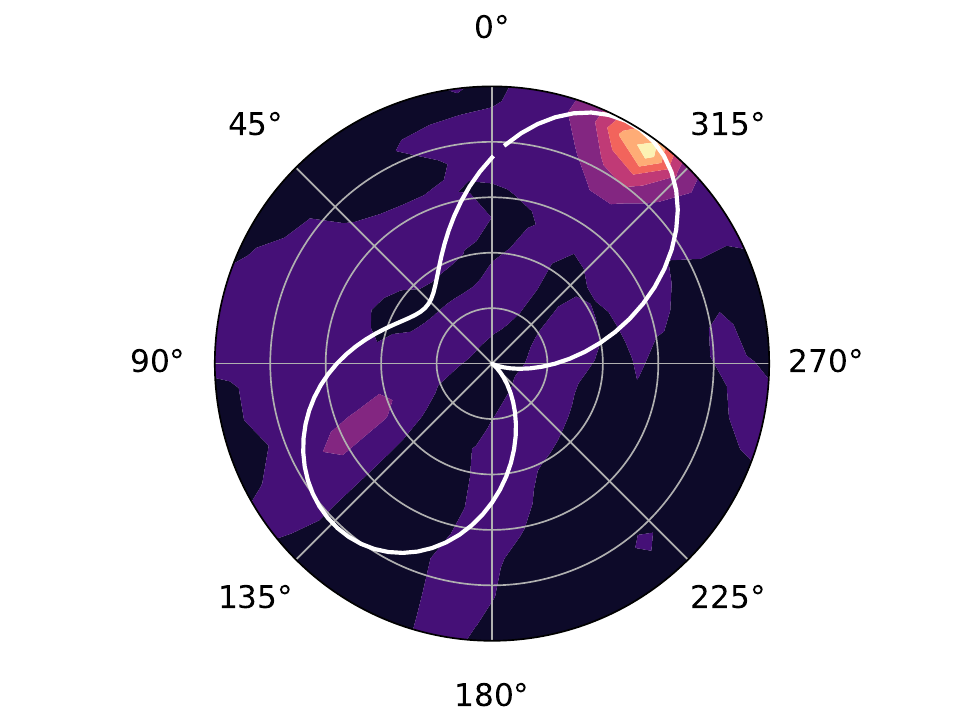}  
  \caption{Segment 4}
\end{subfigure}
\caption{Time-averaged beampatterns (white line) at frequency $f=750$Hz computed on the IcSACC ((a)-(d)) and EcSACC ((e)-(h)) combination weights on 1-second segments from the AMI development set. The heatmap represents the SRP-PHAT acoustic map computed on a 2m radius circular plane. The beampatterns are normalized between 0 and 1.}
\label{fig:avg_bp}
\end{figure*}

This section evaluates the steering directions of both EcSACC and IcSACC on the AMI development set.
Since the speaker locations are unknown, the time-averaged beampattern--Eq. \eqref{fig:avg_bp}--is compared to the SRP-PHAT~\cite{benesty_microphone_2008} energy map.
A position corresponding to a maximum of energy is thought to represent an active acoustic source.
The SRP energy map is compared to the time-averaged beampattern of each complex SACC combination weight.
For visualization purposes, the time-averaged beampatterns are normalized between 0 and 1.

Figure~\ref{fig:avg_bp}~(a)-(d) presents the time-averaged beampatterns of the IcSACC model on 1-second audio segments of the IS1008a session from the AMI development set.
It shows that the main lobe of the beampattern is mostly steered towards the maximum of the energy map.
The main lobe also exhibits significant shape variations. 
In some examples -- figures \ref{fig:avg_bp}(a) and \ref{fig:avg_bp}(c) -- the steering directions are correlated with the source position.
When two sources are active (fig.~\ref{fig:avg_bp}(c)), the model seems to steer towards the directions of both sources.
This model however fails to steer towards the source at 315$^\circ$ as shown in figure \ref{fig:avg_bp}(d).

Figure~\ref{fig:avg_bp}~(e)-(h) presents the time-averaged beampatterns computed on the EcSACC combination weights with the same segments.
The beampatterns exhibit a similar structure as IcSACC with a match between the energy maxima and the main lobe.
In both single- --figures \ref{fig:avg_bp}(a) to (c)-- and dual-sources --figure \ref{fig:avg_bp}(c)-- scenarios, the beampattern correlates with the energy maxima.
On the last segment, the EcSACC model better selects the 315$^\circ$ source direction as shown in figure \ref{fig:avg_bp}(h).

Finally, this section shows that the beampattern is an efficient tool to better interpret the combination weights learned by IcSACC and EcSACC.
This representation highlights the steering directions used by the self-attention modules with a physical interpretation.
The beampattern exhibits a dependence between the combination weights and the source location. 
Based on the AMI data, the model appears to mostly steer toward the acoustic energy maxima.
However, we lack an established protocol to evaluate the steering direction of such systems on a larger scale.
This will be part of a future study.

%% file: 08_invariant.tex
\section{Invariance to the number of microphones}
\label{8_sect_invariance}
The previous sections have shown that multichannel front-end algorithms can be beneficial for distant VAD and OSD.
However, these systems are trained with a fixed array configuration available in the training data. 
Under real meeting conditions, the number of available microphones may vary from one session to another. 
This section proposes a training strategy to enforce the system to generate a unique feature map regardless of the number of microphones provided as input. 
The invariant training is evaluated on the AMI meeting corpus.
This dataset features audio signals captured by two types of arrays located in different positions in the meeting room and allows an evaluation in case of a mismatched array.

\subsection{Channel-number invariant training}

\begin{figure}[t]
    \centering
    \includegraphics[width=\linewidth]{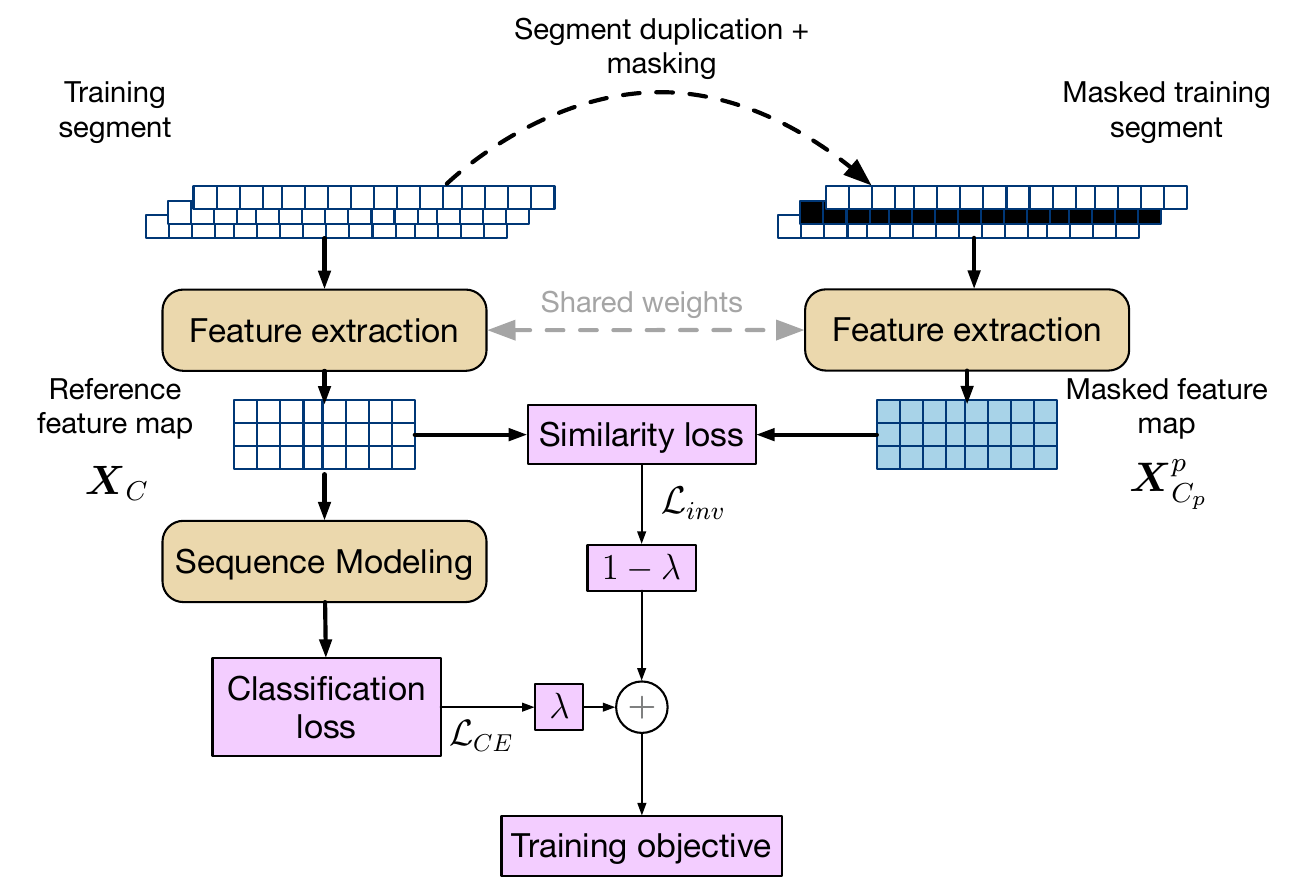}
    \caption{Flowchart of the channel-number invariant training procedure. Only one duplicated segment is represented here ($P$=1).}
    \label{fig:inv_train_principle}
\end{figure}

When a VAD+OSD system is trained with a given microphone array configuration, the performance might be degraded in case of an array mismatch in the testing conditions.
The mismatch can be the result of a change in the microphone array geometry, or a set of microphones stopping to work.
To train a robust system regardless of the number of microphones, the sequence of feature vectors can be enforced to be invariant to the number of channels.
Therefore, we propose an invariant training loss to learn a unique representation of the input signal regardless of the number of active microphones.
The VAD+OSD system is then trained to perform classification and representation invariance in a multi-task formulation jointly.
As detailed in section \ref{sect:2_desc_tasks}, VAD+OSD is formulated as a frame classification task and is optimized using a cross-entropy (CE) training objective $\mathcal{L}_{CE}$. 
An additional training objective, referred to as invariant training loss $\mathcal{L}_{inv}$, is proposed to learn a unique feature map.
The channel-number invariant training is illustrated in figure~\ref{fig:inv_train_principle} and detailed below.
 
Let $g:\boldsymbol{x}_C,\boldsymbol{\Theta} \rightarrow \boldsymbol{X}_C$ be a multi-channel neural feature extractor which maps the raw input signal $\boldsymbol{x}_C$ to a feature map $\boldsymbol{X}_C$. 
$C$ indicates the total number of microphones available in the training data. 
The channel-number invariant training determines a set of parameters $\boldsymbol{\Theta}$ to extract a similar feature vector regardless of the number of active channels.
The training procedure is defined as follows and illustrated in figure~\ref{fig:inv_train_principle}.
The input audio segment $\boldsymbol{x}_C$ is duplicated $P$ times. 
Some channels are then randomly masked in the input and the duplicated segments.
The number of kept active channels $C_p$ is randomly sampled following a uniform distribution $C_p \sim \mathbb{U}_{\{2,C\}}$.
The duplicated masked-segments are collected in a set $\mathcal{D}=\{\boldsymbol{x}^p_{C_p}\}_{p=1}^{P}$.
For each segment in $\mathcal{D}$, the feature map $\boldsymbol{X}^p_{C_p}$ can be computed by feeding the segment $\boldsymbol{x}^p_{C_p}$ to the function $g$. 
We define a channel-number invariant loss, $\mathcal{L}_{inv}$ which forces the masked feature sequence $\boldsymbol{X}^p_{C_p}$ to be as close as the reference feature sequence $\boldsymbol{X}_C$.
This loss minimizes the distance between the reference feature map $\boldsymbol{X}_C$ and each masked representation $\boldsymbol{X}^p_{C_p}$ given the following expression:

\begin{equation}
    \mathcal{L}_{inv}= \frac{1}{P}\sum_{p=1}^P \frac{\|\boldsymbol{X}_{C} - \boldsymbol{X}^p_{C_p}\|_F}{\|\boldsymbol{X}_{C}\|_F\cdot\|\boldsymbol{X}^p_{C_p}\|_F},
    \label{eq:unique_rep_loss}
\end{equation}

\noindent with $\|\cdot\|_F$ being the Frobenius norm. The training objective defined in equation \eqref{eq:unique_rep_loss} aims at generating a unique representation regardless of the number of active channels.
The segmentation model is then trained using the following dual loss:

\begin{equation}
    \mathcal{L} = \lambda \mathcal{L}_{CE} + (1-\lambda) \mathcal{L}_{inv},
    \label{eq:dual_loss}
\end{equation}

\noindent where $\lambda\in [0,1]$ is a trade-off parameter between the two training objectives.

\subsection{Training and evaluation}

The use of invariant training is investigated with our best-performing VAD+OSD system which is the TCN architecture.
Four SACC algorithms are investigated as multi-channel front-ends: SACC+$STFT$, SACC+$\mathcal{A}$, IcSACC, and EcSACC.

The invariant system is trained with audio material from Array 1 of the AMI meeting corpus.
The invariant loss $\mathcal{L}_{\mathrm{inv}}$ is computed by duplicating $P=2$ times the current training segment.
The trade-off parameter between the two loss functions is set to $\lambda=0.7$ since it produces the best OSD performance.

To evaluate the robustness of invariant models, a first evaluation is conducted on signals recorded by the Array~1 by removing a fixed number of channels.
The system is also evaluated using a new array configuration, which the system did not observe during training.
This evaluation is conducted on the data recorded by the Array~2 of the AMI corpus.
It consists of a 4-microphone uniform circular or linear array depending on the session. 
This array is placed on one side of the table, a little further from the participants than the Array~1.
Only OSD performance is presented since it is more representative of the impact of the invariant training.
The performance on both arrays is presented in Table \ref{tab:array1-var_channels}.

\subsection{Evaluation on the AMI Array 1}

This subsection evaluates the impact of the invariant training on VAD+OSD systems. 
The evaluation considers $C=8$, $C=4$, and $C=2$ active channels in the AMI Array 1 evaluation data.
The first part of table \ref{tab:array1-var_channels} shows OSD performance for each configuration.

In the $C=8$ scenario, each kind of system offers the same performance whether the invariant training is considered or not.
For example, SACC+$STFT$ shows a 68.5\% F1-score with the original model and 68.4\% with the invariant training.
Only the EcSACC system degrades by an absolute -1.3\% in the invariant training context.

In the $C=4$ scenario, the performance of the original models degrades with the channel masking.
The SACC+$STFT$ system degrades by -8.5\% between the 8- and the 4-microphone configurations.
The IcSACC and EcSACC models are strongly degraded. 
These systems are not able to detect overlapping speech anymore.
Only the SACC=$\mathcal{A}_{32}$ performance remains stable.
Considering invariant training limits the degradation of the SACC+$STFT$ model (65.9\%).
SACC+$\mathcal{A}_{32}$ retains its performance with the invariant training (65.3\%).
The IcSACC and EcSACC show the best performance with 67.8\% and 67.1\% respectively.

The observations in the $C=2$ scenario are similar to $C=4$.
The original SACC+$STFT$ is degraded by absolute -16.8\%.
This degradation is reduced with the invariant training (64.2\%).
SACC+$\mathcal{A}_{32}$ remains robust to channel dropout (65.3\%) while retaining performance with invariant training (65.6\%).
The invariant training enables IcSACC and EcSACC to detect overlapping speech.
IcSACC offers the best performance with 67.2\%.

\begin{figure}[t]
    \centering
    \includegraphics[width=0.9\linewidth]{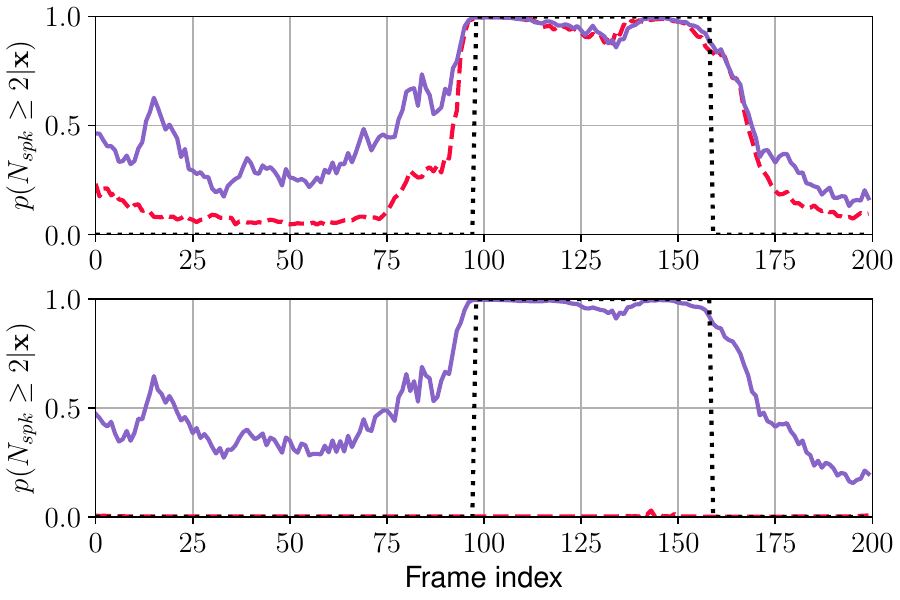}
    \caption{OSD class prediction with the EcSACC model with (Top) $C=8$ microphones and (Bottom) $C=4$ microphones. ($\boldsymbol{\cdots}$) target OSD label, (\textcolor{Orchid}{\textbf{---}}) invariant system prediction, (\textcolor{OrangeRed}{\textbf{-~-~-}}) original system prediction.}
    \label{fig:inv_pred}
\end{figure}

To better understand the loss of information of the original system with missing microphones, Figure \ref{fig:inv_pred} displays the prediction of the EcSACC system on a segment in the $C=8$ and the $C=4$ scenarios by considering invariant training or not.
It shows that the dynamic of the prediction is altered by the loss of microphone pairs and becomes close to zero.
This illustrates the F1-score to be 0.0\%.
Meanwhile, invariant training maintains good prediction in both scenarios.

\subsection{Evaluation on the AMI Array 2}

In the following experiment, each system is evaluated on AMI array~2 data to assess the OSD performance on actual mismatch array configuration.
The OSD performance is presented in the second part of table~\ref{tab:array1-var_channels}. 
In an ideal invariant system, the OSD performance remains consistent from one array to another.
The OSD results show that the original models poorly perform under array mismatch.
For example, the SACC+$STFT$ model reaches a 53.2\% F1-score while it gets 68.5\% on Array 1 with all the channels.
The complex models are unable to detect overlapping speech.
The SACC+$\mathcal{A}_{32}$ system is also degraded (58.9\%) considering the Array 1 performance (64.6\%).

\begin{table}[t]
\centering
\caption{OSD F1-score (\%) on the Array 1 and Array 2 data of AMI evaluation subset. The Array 1 results are computed for a varying number of active channels $C$. Four SACC-based approaches are compared with TCN sequence modeling.}
\begin{tabular}{@{}cclcccc@{}}
\toprule
Array & $C$  &  Model & $STFT$           & $\mathcal{A}_{32}$ & IcSACC & EcSACC\\ \midrule
\multirow{6}{*}{1}&\multirow{2}{*}{8}  & Original      & 68.5 & 64.6 & 67.6 & 68.4\\
&                    & \quad /w $\mathcal{L}_{inv}$    & 68.4 & 65.2 & 67.2 & 67.1\\ 
\cmidrule{2-7}
&\multirow{2}{*}{4}  & Original                        & 60.0 & 65.2 & 0.0 & 0.0\\
&                    & \hfill /w $\mathcal{L}_{inv}$   & 65.9 & 65.3 & 67.8 & 67.1\\ 
\cmidrule{2-7}
&\multirow{2}{*}{2}  & Original                        & 51.7 & 65.3 & 0.0 & 0.0\\ 
&                    & \hfill /w $\mathcal{L}_{inv}$   & 64.2 & 65.6 & 67.2 & 66.7\\ 
\midrule
\multirow{2}{*}{2}& \multirow{2}{*}{4}  & Original     & 53.2 & 58.9 & 0.41 & 0.0\\
                & & \quad /w $\mathcal{L}_{inv}$       & 63.4 & 57.6 & 59.6 & 56.5\\ 
\bottomrule
\end{tabular}
\label{tab:array1-var_channels}
\end{table}

Adding the invariant training drastically improves the robustness of most systems.
The SACC+$STFT$ features a +10.2\% absolute improvement considering the original model with a 63.4\% F1-score.
Both IcSACC and IcSACC can detect overlapping speech with 59.6\% and 56.5\% F1-score respectively.
Only the SACC+$\mathcal{A}_{32}$ system shows similar performance with 57.6\%.
The experiments on Array 1 data show that this model highlights some robustness to channel masking.
This is confirmed under mismatched array conditions.

In this section, we have proposed a new training procedure to learn a similar feature sequence regardless of the number of input microphones.
The feature sequence invariance is learned at the same time as classification in the multi-task learning framework.
This section shows that SACC-based VAD+OSD performance tends to be degraded under array configuration mismatch.
To tackle this issue, a channel-number invariant loss $\mathcal{L}_{inv}$ is proposed.
AMI Array~1 experiments confirm the suitability of the invariant training framework for OSD, mostly with IcSACC and EcSACC.
Under invariant training conditions, the systems can keep the OSD performance close to the $C=8$ scenario, i.e. when all the channels are active.
On Array 2 data, the performance is slightly degraded, but still improved concerning the original models.

%% file: 09_conclu.tex
\section{Conclusions and perspectives}
\label{9_sect_conclusion}

This paper presents several multi-channel front-ends for joint distant Voice Activity and Overlapped Speech Detection (VAD+OSD) for speaker diarization. 
These algorithms are all based on the same principle: weighting and combining the channels coming from a microphone array.
Three combination-weight estimation procedures were proposed.
Those are all inspired by the Self-Attention Channel Combinator (SACC), which estimates combination weights from the multi-channel short-time Fourier transform (STFT) magnitude.

In the first instance, the STFT is replaced with a learnable filter bank based on analytical filters.
The other two methods exploit the magnitude and phase of the STFT in explicit (EcSACC) and implicit (IcSACC) ways.
Each approach is investigated as a front-end of both BLSTM- and TCN-based VAD+OSD systems.
The self-attention-based models achieve similar performance as the MVDR without requiring the costly estimation of the covariance matrices.
Among the SACC extensions, the learnable filter bank exhibits mitigated OSD results while complex extensions drastically improve the performance in the distant speech scenario e.g. EcSACC achieves 68.4\% F1-score on the AMI evaluation set with the TCN model.
On the VAD task, the segmentation error rate (SER) can be improved with channel-combination algorithms.

The impact of VAD and OSD is evaluated the final back-end task, i.e. speaker diarization.
Complex SACC extensions offer the best diarization performance with 23.30\% DER and 30.90\% JER for EcSACC.
The assignment of overlap segments highlights the need for robust OSD since one can expect up to +36.3\% relative DER improvement, as shown by Oracle's performance.

The complex SACC extensions achieve competitive VAD+OSD and speaker diarization performance.
These approaches are also designed to be more explainable.
The analysis of the combination weights with the beampattern exhibits a correlation between the maxima of acoustic energy and the steering directions.
The EcSACC seems better at steering towards the source.
Finally, the performance of the proposed front-ends highly relies on the microphone array configuration used during training.
Thus, a mismatch between the training and the testing array setup may lead to severe performance degradation.
To minimize this degradation, a new loss function is proposed and added to the training process.
This training objective consists of learning a unique feature sequence regardless of the number of available microphones.
Experiments conducted on a TCN-based VAD+OSD system with four front-ends demonstrate that the performance remains steady regardless of the number of active microphones (SACC+$STFT$: +10.2\% F1-score improvement on the mismatched data from AMI array~2). 
In the case of EcSACC and IcSACC, this loss prevents the model from failing the detection.

In future work, new training objectives to learn a channel-invariant representation will also be investigated.
The evaluation of invariant models in the cross-corpus scenario will also be considered.
Since the complex combination weights encode spatial information about the active source, they could bring more information to the OSD system.
Complex combination weights will thus be investigated as additional spatial features.


